\documentstyle{article}
\makeindex

\def\esp#1{e^{\,\textstyle#1}}

\def\relaz#1#2{\mathrel{\mathop{\kern 0pt#1}\limits_{#2}}}

\def\H{{\cal H}}
\def\L{{\cal L}}

\begin{document}

\title{ \Large \bf  Effective Potential and Quantum Dynamical Correlators.}
\author
{\large \bf R. Giachetti$^{1,2}$, R. Maciocco $^{1,2,3}$ and V.
Tognetti$^{1,3}$
\\
 $^1$Dipartimento di Fisica, Universit\'a di Firenze\\
${^2}$ Istituto Nazionale di Fisica Nucleare, Sezione INFN di
Firenze,\\ ${^3}$ Istituto Nazionale di Fisica della Materia,
Unit\'a INFM di Firenze}

\date{\today}

\maketitle

\baselineskip=13pt
{\footnotesize The approach to the calculation of quantum dynamical
correlation functions is presented in the framework of the Mori
theory. An unified treatment of classic and quantum dynamics is
given in terms of Weyl representation of operators and holomorphic
variables. The range of validity of an approximate quantum
molecular dynamics is discussed.}

\bigskip
PACS N.: 05.30.-d, 05.60.+w, 67.20.+k
\bigskip
\baselineskip=19.25pt
\bigskip

Methods that permit to determine, at least approximately, the
thermodynamical properties of quantum dynamical systems by means of
calculations of classical type usually turn out to be very useful. For
the static case, this goal has been successfully achieved by means of
a variational procedure based on path integral formulation of
statistical mechanics \cite{Feystat}. The method leads to the
construction of an effective potential that accounts completely for
the quantum corrections of the quadratic part of the interaction (the
{\it harmonic oscillators}) and treats in a self-consistent one loop
approximation the quantum effects due to nonlinear terms
\cite{GTall,FeynmanK86,CowleyH95}.  The effective potential has proved
to be a very powerful tool for concrete computations
\cite{CTVV92,CTVVmagall,CGTVV95}: in particular it has allowed for the
reduction of quantum Monte Carlo to classical Monte Carlo numerical
calculations \cite{CGTVV95}, with evident fruitful consequences
\cite{CMGTVall}.

A rigorous treatment of the dynamical problem is, as always, much more
difficult. A success in such a program, however, is scientifically
very remunerative. Indeed, for  instance, it opens the way to a sound
formulation of molecular dynamics for quantum systems. Using the
effective potential obtained in \cite{GTall,FeynmanK86}, some naive
attempts of extending the treatment to a dynamical setting have been
tried \cite{CaoVall}. However, by admission of the authors themselves,
these attempts assume some {\it ad hoc} multiplicative relation in
order to connect the ``centroid molecular dynamics" to the quantum
correlation function \cite{CaoVall,MartynaCao}.

In this paper we shall present a new more rigorous approach, based on
the Mori theory and on projection techniques \cite{Moriboth}. At
lowest order, but within the framework of a consistent analysis, we
recover the heuristic expressions of ref. \cite{CaoVall} showing their
limitations and indicating the correct method to improve them.

The key points of our theory are: $i)$ the dynamics of classical and
quantum oscillators is the same; the only difference is the static
amplitude factor determined by the initial conditions. $ii)$ the use
of the Weyl representation \cite{Weyl} which translates quantum
problems into a classical framework and permits a unified approach by
which any semiclassic expansion is obtained in natural way
\cite{Faddeev}.

Starting from what we have learned in the theory of the effective
potential \cite{CTVV92,CGTVV95}, we develop a semiclassical
calculation in the Weyl representation \cite{CTVV92}, that describes
quantum mechanics in terms of the classical canonical variables $q$
and $p$,  expressed in terms of the holomorphic variable
$z=2^{-1/2}(q-ip)$, $z^*=2^{-1/2}(q+ip)$, corresponding to the
creation and  annihilation operators $a^\dag$ and $a$. If we consider
a dynamical system with a Hamiltonian $H$, the quantum equation of
motion of an operator $A$, in the Weyl representation, reads
\cite{Turco}
\begin{equation}
\frac{dA^w}{dt}=\frac
2\hbar\,\H^w\,\sin\,\frac{\hbar\Lambda}2\, A^w=i\L A^w.
\label{Eq.Moto}
\end{equation}
Here $\Lambda$ is the differential operator that gives the Poisson
bracket,
\begin{equation}
\Lambda\,=\,i\,[{\overleftarrow \partial}_z\, {\overrightarrow \partial}_{z^*}\,-\,
{\overleftarrow \partial}_{z^*}\, {\overrightarrow \partial}_z]\,.
\end{equation}
and $\L$ is called the Liouville operator. $A^w$ and $\H^w$ are the
Weyl symbols of the corresponding quantum observables. Explicitly
\begin{equation}
{\cal O}^w(z^*,z)= \esp{-zz^*}\int \frac{d\zeta d\zeta^*}{\pi
i}\esp{(-\zeta^*\zeta+z^*\zeta-\zeta^*z)}\,\langle z-\zeta|{\cal
O}|z+\zeta\rangle,\label{Weyl}
\end{equation}
where $|z\rangle$ is the coherent state, defined as:
\begin{equation}
|z\rangle=\esp{\frac12 z^*z}\,\esp{z^*a^\dag-za}\,|0\rangle.
\end{equation}
The canonical equations are obviously recovered when $\hbar\rightarrow
0$. Moreover the statistical average of a dynamical variable reads
\begin{equation}
\langle{\cal O}
\rangle={\rm Tr}\,(\rho\,{\cal O})=\int  \frac{d\zeta d\zeta^*}{2\pi i}\,
\rho^w(z^*,z)\,{\cal O}^w(z^*,z)\label{Average}
\end{equation}
and, again, the  classical limit reproduces the phase-space average.

For sake of simplicity, we shall only consider a system with one
degree of freedom and with a Hamiltonian
\begin{equation}
\H=\hbar \omega_0\,\, z^*z+gV(z^*,z),\label{Hamiltonian}
\end{equation}
where  $g$ is the quantum coupling. In the following, when not
necessary, we put $\hbar=1$. The equation of motion for $z$ and $z^*$
can be put into the following form along the line of the projection-
Zwanzig-Mori theory \cite{Moriboth}:
\begin{equation}
\frac{dz(t)}{dt}=i\omega z(t)+\int_0^td\tau\,M(\tau)z(t-\tau) +
f^w(t)\,.\label{Langevin}
\end{equation}
A similar equation can be written for $z^*(t)$, that, as a quantum
observable, must be treated independently of $z$.

This equation can be solved once the initial condition $z(0)$ is
assigned. In the classical context one has to  fix the values
$q(0)$ and $p(0)$, while in the quantum case the initial state of
the Hamiltonian must be specified({\ref{Hamiltonian}). In the
following we shall use the shorthand notation $z(0)\equiv z$ and
the analogous for other dynamical variables.

The term $f(t)$ describes the effect of the fluctuating forces, caused
by the non linear interaction $gV(z^*,z)$ and has the following
expression:
\begin{equation}
 f^w(t)
=\esp{it(1-P)\L}\,(1-P)\L\,z\,,\label{Forza}
\end{equation}
where $P$ is the projection operator onto the direction of $z$. The
explicit expression of $P$, as well as the memory kernel $M(t)$ are
conveniently given in terms of a scalar product introduced by Mori
\cite{Moriboth}. The latter plays an important role in the theory and,
for quantities with vanishing thermal averages, reads:
\begin{equation}
(A|B(t)) =
\int_0^\beta\,d\lambda\,\langle\,{(A^w)}^*(0)\,B^w(t+ i\lambda)\,\rangle\,.\label{Mori}
\end{equation}
It turns out that:
\begin{equation}
P_{A}\,B(t)= (A|B)^{-1}(A|B(t)).\label{P}
\end{equation}

\noindent
One realizes that this is just the Kubo function and in the classical
limit reduces to the correlation function up to a multiplicative
factor.

The quantities of physical interest are the Fourier transforms
$C(\omega)$ and $R(\omega)$ of the correlation function,$\,
\langle a^\dag a(t)\rangle\,$-- probed by scattering experiments  -- and Mori
product, $\,(a|a(t))\,$ respectively. They are connected by the
``detailed balance" principle:
\begin{equation}
C(\omega) =
\frac {\omega}{1-\esp{-\beta\omega}}R(\omega).\label{Dettaglio}
\end{equation}

The frequency $\omega$ is given by:
\begin{equation}
i\omega=(z|z)^{-1}\,(z|P\L z)\,.\label{freq}
\end{equation}
while the memory kernel $M(t)$ turns out to be
\begin{equation}
M(t)=(z|z)^{-1}\,(f^w(t)|f^w)\,,\label{Memory}
\end{equation}
with $f^w=f^w(0)$. Referring to eq.(\ref{Langevin}), $M(\tau)$ can be
written in the form
\begin{equation}
M(\tau)=\delta_1\,\Xi_1\,,~~~~~~~~~~\Xi_1=(f_1^w|f_1^w)^{-1}
(f_1^w(\tau)|f_1^w)\,\label{Kernel}
\end{equation}
where $~f^w_0= z$, $~\omega_0= \omega$, $~P_0= P$, $~\L_ {(0)}=
\L~$ and
\begin{equation}
\delta_1= (z|z)^{-1}(f_1^w|f_1^w).\label{Normal}
\end{equation}
It will be necessary to introduce some further definitions in order to
describe the recursive framework of the dynamical theory.  The new
quantities at different orders will be denoted by different indices.
We want, however, to stress that the factor $\delta_1$ is defined only
in terms of static quantities.

The equation of motion for
\begin{equation}
\Xi_0(\tau)=(z|z)^{-1}( z|z(\tau))\,.\label{Xi}
\end{equation}
becomes:
\begin{equation}
\frac{d\Xi_0}{dt}=i\omega\,\Xi_0(t)+\delta_1\,
\int_0^td\tau\,\Xi_1(\tau)\,\Xi_0(t-\tau). \label{XiXi}
\end{equation}

Denoting by $P_j$ a family of projection operators in the direction of
the vectors $f_{j}$, from (\ref{XiXi}) we easily generate a hierarchy
of equations with  reduced Liouvillians:
\begin{equation}
{\cal L}_j=(1-P_j-P_{j-1}-.....-P_1-P_0)\L_{0}\,.\label{Elle}
\end{equation}
Finally, the continued-fraction expansion for the Laplace transform
$\Xi_0(z)=R(z)/R(t=0)$ is obtained \cite{Moriboth}:
\begin{equation}
\Xi_j(z)={{1}\over{z+i\omega_j+\delta_{j+1}\Xi_{j+1}(z)}}\,,\:\:\:\omega_j=\frac
{(f^w_j|\L_jf^w_j)}{(f^w_j|f^w_j)}\,,\:\:\:
\delta_{j+1}={{(f^w_{j+1}|f^w_{j+1})}\over{(f^w_{j}|f^w_{j})}}\,.\label{Continue}
\end{equation}

The quantities $\omega_j$ and $\delta_{j+1}$ are related to the
first $2(j+1)$ moments of the time series expansion of $\Xi_0(t)$.
While $(f^w_{j}|f^w_{j})$ with $j\not= 0$ can be expressed in terms
of static correlation, the normalization quantity
$(f^w_{0}|f^w_{0})$ requires the direct evaluation of the Mori
product, giving the static susceptibility.

We observe that the calculation of $\Xi(z)$ -- and therefore of
$C(\omega)$ -- can be approached from the knowledge of the static
quantities $\delta_j$ up to a sufficiently large number $j=J$ ,
supported by some insight into the long time behaviour of the
continued fraction termination $\Xi_J(t)$ \cite{LoveseyM72} . The
effective potential theory permits its approximate determination
\cite{CGTVV95} and  recently the averages of the imaginary time
ordered products of operators that involve both momentum and
displacement have been evaluated \cite{CGTVprl?}.

Let us consider  the parameter $f=\beta\hbar \omega/2$, introduced
by Feynman, which determines the quantum character of the system
through the Langevin function: $L(f)=(\coth\,f - 1/f)\,$. From the
results of  \cite{CGTVprl?}, we derive explicit expressions for the
average of an operator ${\cal O}$
\begin{equation}
\langle {\cal O}\rangle= \frac 1{\cal Z}\int \, \frac{dz^*dz}{2\pi i} \,\langle\langle
{\cal O}^w(z^*, z)\rangle\rangle \,\esp{-\beta {\cal
H}_{eff}(z^*,z)}\,,\label{genav}
\end{equation}
and for the Mori product, at lowest order in $f$:
\begin{equation}
({\cal A}|{\cal B})\,\sim\,\frac 1{\cal Z}\beta\int
\,\frac{dz^*dz}{2\pi i}
\,\langle\langle {\cal A}^w(z^*,z)\rangle\rangle\,
\langle\langle {\cal B}^w(z^*,z)\rangle\rangle\,\esp{-\beta {\cal
H}_{eff}(z^*,z)}\,,\label{gemori}
\end{equation}
where ${\cal H}_{eff}$ is the effective Hamiltonian,
$\,\langle\langle {\cal O}^w(z^*,z)\,\rangle\rangle$ is the
two-dimensional gaussian average of the Weyl symbol of the operator
${\cal O}(p,q) $\cite{CTVV92,CGTVV95} and
\begin{equation}
{\cal Z}=\int \, \frac{dz^*dz}{2\pi i} \,
\,\esp{-\beta {\cal H}_{eff}(z^*,z)}\,,
\end{equation}
is the partition function.

For instance, by this expression, (\ref{genav}) the static average
of a quantity $A(q)$ dependent on the coordinates only, turns out
to be
\begin{equation}
\langle  A \rangle=\frac 1 {\cal Z}
\sqrt{\frac m{2\pi\beta\hbar^2}}\,\int d\xi
\langle\langle A(\xi)\rangle\rangle\esp{-\beta V_{eff}(\xi)}\,,
\label{statA}
\end{equation}
where
\begin{equation}
\langle\langle A(\xi)\rangle\rangle\,=\,(2\alpha \pi)^{-1}\int d\eta
A(\xi+\eta)\,\esp{-\eta^2/(2\alpha)}\,\label{statAA}
\end{equation}
denotes the gaussian average of the variable $A(x)$ with moment
$\,\alpha=\hbar/(2m\omega)\,L(f)\,$.

For $A=q^2$ the relation (\ref{statA}) gives $\langle q^2\rangle
=\alpha+\hbar
/(2mf\omega)$, where the first term $\alpha$ is a consequence of eq.(\ref{statAA}),
while the second term is due to the "classical" average with the
effective potential. This means that the gaussian average described
in eq. (\ref{statAA}) is essential to yield the correct quantum
occupation number $\langle a^\dag a\rangle$ from the classical one.
This is a crucial point for a good approximation of quantum
dynamics.

A perturbative approach of the Mori theory  can be performed when
the Liouvillian  ${\cal L}$ can be split in two terms ${\cal L}^d$
and ${\cal L}^{nd}$ satisfying the relations:
\begin{equation}
P_0\L P_0=\L^d\,,\,\,\,\,\,\,(1-P_0)\L P_0
=\L^{nd}\,\label{Pertur}
\end{equation}
i.e ${\cal L}^d$ cannot connect diagonal with non-diagonal terms
and viceversa, while ${\cal L}^{nd}$ makes always this connection.
In an exact splitting of the operator $\L$ we should extract all
diagonal contributions from the non linear terms. This can be done,
by a one loop approximation, which causes renormalization of
frequency and coupling. This renormalization is caused both by
quantum and thermal gaussian fluctuations.

Here, we want to propose a different perturbative scheme, in the
spirit of the effective potential method. We account exactly for
the classical part as well as for the fully quantum contribution of
the harmonic oscillators. The purely quantum effects of the
nonlinear terms will then be calculated in a self consistent way at
the level of one loop approximation. In particular, when operating
on the nonlinear part $g\H^w_{nl}$, at lowest order in the non
linear quantum coupling we shall be allowed for the following
substitution:
\begin{equation}
(2/\hbar)\,g{\cal H}^w_{nl}\,\sin\,(\hbar\Lambda/2)\,
\sim\,g{\cal H}^w_{nl}\,\Lambda.\label{Approx}
\end{equation}

Accordingly, we split the Liouvillian operator into the sum
$\L=\L^h+g\L^{an}$, where
\begin{equation}
\L^h=2\Omega\,(z^*z+\frac
12)\,\sin\,\frac{\hbar\Lambda}2\,, ~~~~~~~~~~
g\L^{an}=g\,W(z^*z)\,\Lambda\,\label{splitting}
\end{equation}
The parameters $\Omega$ and $W$ will be determined by physical
requirements. It is apparent from Eq.(\ref{splitting}) that the
quadratic part is treated in fully quantum way,
 giving the quantum evolution of the
``best" harmonic oscillator, while the non linearity causes a
``classical-like" dynamics. As a consequence of the splitting
$\L=\L^h+\L^{an}$, at the lowest nonvanishing order in the coupling
constant,   we have these important consequences. The quantum
effects on dynamics refers to the imaginary part of the path
($0,i\beta$). We take into account that the parameters of the
effective potential are calculated at the average point of this
path along which it is affected by an harmonic potential.  As a
consequence, $\Omega$ and $W$ are exactly the same as defined by
the effective potential for the static case.

The validity of the use of the effective potential is related to the
quantum coupling $g$. At the first order, corresponding to a range of
temperatures,  $\beta<1/g$, the definition (\ref{Forza}) of
$f_{1}^w(i\lambda)$ becomes:
\begin{equation}
f_{1}^w(i\lambda)\,\sim\,\esp{-\lambda(1-P_0){\cal
L}^h}\,f_{1}^w\,=\,\esp{-\lambda{\cal
L}^h}\,(1-P_0)g\L^{an}\,z.\label{Fappr}
\end{equation}

With the approximate expression (\ref{Fappr}) of the fluctuating
force, the integration on $\lambda$ in the relaxation function can be
easily performed. The result of this integration, is cancelled out by
the denominator in the expressions of $\Xi_1$, so that only
correlation functions done by the rules given by eqs. (\ref{genav}),
(\ref{statA}) are involved and the parameter $\delta_1$ is calculated
with the same receipt. The procedure can be iterated at successive
levels, using again the approximate Liouvillian of
eq.(\ref{splitting}). Therefore a continued fraction expansion is
obtained where each coefficient $\delta$'s is calculated within the
afore-mentioned scheme. However it is worthwhile to notice that the
nonlinear quantum effects on $\delta$'s are evaluated only at one
loop, so that their inaccuracy is increasing with the order of
$\delta_n$.

Furthermore, we can do a rather crude approximation for the memory
function $M(t)$. Let us neglect the Gaussian quantum spread
(\ref{statAA}) of the product of the fluctuating force, under the
condition  $(f_{1}(t)f_{1})^w\,\sim\,f_{1}^w(t)f_{1}^w $. This
corresponds to neglecting the quantum effects of non commutability
of the operators in the expression of the memory function and using
the classical expressions for occupation numbers.

At this level the equation of motion for $z(t)$ becomes
\begin{equation}
\frac{dz(t)}{dt}=i\bar{\Omega}\,z(t)+\delta_1\,
\int_0^td\tau\,\Xi_1(\tau)\,z(t-\tau) + f_{1}^w(t)\,,\label{class}
\end{equation}
where $\bar{\Omega}$ is the frequency $\Omega$ renormalized by the
"diagonal" part of the classical non-linearity\cite{Moriboth},
containing only correlation functions, (see Eq.\ref{freq}). At the
same level, $M(\tau)=\delta_1\Xi_1$, is given simply by a
classical-like correlation function. Therefore eq.(\ref{class})
represents a "classical" evolution of $z(t)$ where the frequency
$\Omega$ and the potential $W$ must be determined in such a way to
reproduce the same static average values of the observables. At
this approximation level, therefore, they must coincide with the
quantum-renormalized frequency and with the effective potential
parameter of the static case \cite{CTVV92}. At the same level the
Mori product $\Xi_0(t)$ is nothing but thedynamical correlation
function calculated by the classical-like effective potential.
Finally, through the detailed balance principle, we can recover the
spectral shape as guessed in ref. \cite{CaoVall}.

Of course this approach is valid for $g\rightarrow 0$, $i.e.$
classical case or non interacting harmonic oscillators. As the
coupling $g$ increases, its validity is limited to short times. The
range is dependent, firstly on the smallness of the parameter $f$,
because we have neglected the gaussian quantum spread of the
fluctuating forces, secondly on the inverse of the quantum coupling
$g$.  Indeed, only the normalized second moment $\delta_1\,$ is rather
well reproduced. All effects related to line shape, like finite
lifetimes of the elementary  excitations are considered at classical
level so that there is a shift of the peak of the lineshape, without
any substantial modification of the linewidth, \cite{PCMTV98}.

A  further improvement, proposed by \cite{CaoVall}, can be
suggested, from an heuristic point of view,  by the lowest order
expression for the Mori product in the memory function,
\cite{CGTVprl?} shown in eq.(\ref{gemori}). At this order the Mori
product can be reduced to the average, done by the effective
potential, of the product of the gaussian quantum averages of the
fluctuating forces. This observation could justified the insertion
of the gaussian quantum average of the forces in the "centroid
molecular dynamics" (\cite{CaoVall,MartynaCao}).

Our derivation shows the limitations of a ``quantum molecular
dynamics" done by means of the simple use of the effective
potential in the classical equations of motion. The simple
corrections of the occupation numbers, recently proposed
\cite{CaoVall,MartynaCao}, cannot satisfy the request of the
calculation of the dynamical correlations for quantum-interacting
(strong correlated) systems. The necessity to account for quantum
effects in the life-time of the excitation would require anharmonic
corrections to effective potential on the same kind of the ones
proposed in heuristic way \cite{CowleyH95,ACHprl}. Nevertheless the
excellent results  for the thermodynamic quantities \cite{ACHprl}
this correction has not been completely justified up to now  nor
its extention to the static quantities appears to be
straightforward. The quantum computations of dynamical correlations
is still an open problem which has to be faced by different
approaches. The moment calculations and the consequent expansions
\cite{CGTVV95,CMGTVall} although not exhaustive, are still
competitive.

\end{document}